\documentclass[preprint,eqsecnum,aps,nofootinbib]{revtex4}
\usepackage{amsfonts,amsmath,amssymb,amsthm}
\usepackage{latexsym}
\usepackage{bbm,bm}
\usepackage{graphicx}

\newcommand{\ket}[1]{\lvert #1 \rangle}

\newcommand{\beq}{\begin{equation}}
\newcommand{\eeq}{\end{equation}}
\newcommand{\beqs}{\begin{eqnarray}}
\newcommand{\eeqs}{\end{eqnarray}}

\begin{document}

\title{Is entanglement a unique resource in quantum illumination?}

\author{MuSeong Kim$^1$, Mi-Ra Hwang$^1$, Eylee Jung$^1$, and DaeKil Park$^{1,2}$\footnote{corresponding author, dkpark@kyungnam.ac.kr} }

\affiliation{$^1$Department of Electronic Engineering, Kyungnam University, Changwon,
                 631-701, Korea    \\
             $^2$Department of Physics, Kyungnam University, Changwon,
                  631-701, Korea }

%

\begin{abstract}
It is well-known that quantum illumination with a two-mode squeezed vacuum state as an initial entangled bipartite state achieves $6$ dB quantum advantage in the error probability compared
to classical coherent-state illumination. Is entanglement the only resource responsible for the quantum advantage? We explore this question by making use of squeezing operations. 
Finally, we conclude that the answer to the question is negative. 
\end{abstract}
\maketitle

\section{Introduction}
Quantum entanglement\cite{schrodinger-35,text,horodecki09} is known to be a physical resource in the various types of quantum information processing (QIP).
It is used in  many QIP such as  in quantum teleportation\cite{teleportation,Luo2019},
superdense coding\cite{superdense}, quantum cloning\cite{clon}, quantum cryptography\cite{cryptography,cryptography2}, quantum
metrology\cite{metro17}, and quantum computers\cite{qcreview,computer,supremacy-1}. Quantum computing in particular  attracted a lot of  attention recently after IBM and Google both independently created quantum computers.
It is debatable whether  ``quantum supremacy'' is achieved or not in the quantum computation. 


A few years ago another type of  entanglement-assisted QIP called quantum illumination\cite{lloyd08,tan08} became of interest to the research community. The purpose of this protocol is to detect  low reflective objects embedded in baths of  strong thermal noise.
Typical quantum illumination is described in the  following. The transmitter generates two entangled photons known as the signal (S) and idler (I) modes. The S mode photon is used to interrogate 
an unknown object hidden in the background. After receiving a photon from the target region, joint quantum measurement of the returned beam from target region and the retained I-mode photon is performed to decide on the absence or presence of a target.
In particular, authors in Ref. \cite{tan08} used the two-mode squeezed vacuum (TMSV) state as an initial entangled bipartite state between S and I modes.
The most surprising result from the quantum illumination is the fact that the error probability related to detection is drastically lowered compared to classical coherent-state illumination, even if the initial entanglement between S and I modes disappears 
due to strong background noise. In particular, Ref.\cite{tan08} obtains a quantum advantage of $10 \log_{10} 4 \approx 6.02$ dB in terms of the error probability when compared to classical  illumination. 
An experimental realization of quantum illumination was explored in Ref. \cite{guha09,lopaeva13,barzanjeh15,zhang15,zhuang17}.

Quantum illumination with  Gaussian states has been extended to asymmetric Gaussian hypothesis testing\cite{asymmetry-1,asymmetry-2}. 
Also, quantum illumination with non-Gaussian initial states generated by photon subtraction and addition has also been discussed \cite{zhang14,fan18}.
More recently, quantum illumination with three-mode Gaussian states was examined\cite{eylee21}. 
Another important issue in quantum illumination is the need to develop a quantum receiver, where joint measurement of the returned and retained I mode beams can be performed.
There have been several proposals related to quantum receivers\cite{guha09,guha09-2,zhuang17,Jo21-2} and even their demonstrations\cite{zhang15,bar20}.

Is entanglement a unique resource responsible for quantum advantage in the quantum illumination? In this paper we explore this question by making use of squeezing operations.
In Sec. II we briefly review Ref. \cite{tan08}. 
In Sec. III we apply two single-mode squeezing operations to the TMSV state. Since the two single-mode operations are local unitary, it is obvious that the resulting state 
has the same entanglement with the TMSV state. Nonetheless, it is shown that the quantum advantage in the error probability reduces with increasing the squeezing parameter, and 
eventually the quantum disadvantage occurs when the squeezing parameter is larger than some critical value. In Sec. IV we apply two-mode squeezing operations to the TMSV state. 
It is shown that the resulting state has larger entanglement than the TMSV state. In spite of larger entanglement the quantum advantage decreases with increasing the squeezing parameter. 
From the results of Sec. III and Sec. IV we conclude that entanglement is not the only resource responsible for the quantum advantage in quantum illumination, which is summarized in Sec. V.

\section{Brief Review of Two-mode Gaussian quantum illumination}
The authors of Ref.\cite{tan08} used a TMSV state as the initial bipartite state of the S and I modes in the form;
\begin{equation}
\label{two-signal-state}
\ket{\psi}_{SI} = \sum_{n=0}^{\infty} \sqrt{\frac{N_S^n}{(1 + N_S)^{n+1}}} \ket{n}_S \ket{n}_I.
\end{equation} 
This is a zero-mean Gaussian state whose covariance matrix is 
\begin{eqnarray}
\label{two-signal-state-2}
V_{TMSV} = \left(     \begin{array}{cccc}
                           A  &  0  &  C  &  0            \\
                           0  &  A  &  0  &  -C           \\
                           C  &  0  &  A  &  0            \\
                           0  &  -C  &  0  &  A
                                \end{array}               \right)
\end{eqnarray}
where $A = 2 N_S + 1$ and $C = 2 \sqrt{N_S (1 + N_S)}$. 

Let $\rho_0$ and $\rho_1$ be the bipartite quantum states of the returned beam from the target region and the retained I-mode photon when a target is absent and present, respectively. 
Both are zero-mean Gaussian states. Since, for $\rho_0$, the annihilation operator 
for the return from the target region will be $\hat{a}_R = \hat{a}_B$, where $\hat{a}_B$ is the annihilation operator for the thermal state that has an average photon number $N_B$, its covariance matrix 
can be written in the form:
\begin{eqnarray}
\label{TMSV-0}
V_0 =   \left(      \begin{array}{cccc}
                       B  &  0  &  0  &  0                   \\
                       0  &  B  &  0  &  0                   \\
                       0  &  0  &  A  &  0                   \\
                       0  &  0  &  0  &  A
                         \end{array}                       \right)
\end{eqnarray}
with $B = 1 + 2 N_B$. 
For $\rho_1$ the return-mode's annihilation operator would be $\hat{a}_R = \sqrt{\kappa} \hat{a}_S + \sqrt{1 - \kappa} \hat{a}_B$, 
where $\kappa$ is the reflectivity from a target and  $\hat{a}_B$ is the annihilation operator for a thermal state with an average photon number $N_B / (1 - \kappa)$. We assume a very lossy $(\kappa \ll 1)$ return from a target
with a strong thermal background $(N_B \gg 1)$. Then, the covariance matrix of the $\rho_1$ can be written in the form
\begin{eqnarray}
\label{TMSV-1}
V_1 = \left(      \begin{array}{cccc}
                 F  &  0  &  \sqrt{\kappa} C  &  0                   \\
                 0  &  F  &  0  &  -\sqrt{\kappa} C                  \\
                 \sqrt{\kappa} C  &  0  &  A  &  0                   \\
                 0  &  -\sqrt{\kappa} C  &  0  &  A
                      \end{array}                                            \right)
\end{eqnarray}
where $F = 2 \kappa N_S+ B$.

In order to accomplish  quantum illumination processing, hypothesis testing should be performed to determine  whether or not a target is present.
We take the null hypothesis $H_0$ to mean target absence and the alternative hypothesis $H_1$ to indicate target presence. 
Then, the average error probability is 
\begin{equation}
\label{error-1}
P_{E} = P(H_0) P (H_1 | H_0) + P(H_1) P(H_0 | H_1)
\end{equation}
where $P(H_0)$ and $P(H_1)$ are the prior probabilities associated with the two hypotheses. We assume $P(H_0) = P(H_1) = 1 / 2$ for simplicity.
The two kinds of errors $P (H_1 | H_0)$ and $P(H_0 | H_1)$ are usually referred to as type-I (false alarm) and type-II (missed detection) errors, respectively.
Therefore, the minimization of $P_E$ naturally requires optimal discrimination of $\rho_0$ and $\rho_1$. 

If we have $M$ identical copies of $\rho_0$ and $\rho_1$, the optimal discrimination scheme presented in Ref. \cite{sacchi05-1,sacchi05-2} yields the minimal error probability $P_E^{min}$ in the form
\begin{equation}
\label{error-2}
P_E^{min} = \frac{1}{2} \left[ 1 - \frac{1}{2} || \rho_0^{\otimes M} - \rho_1^{\otimes M} ||_1 \right]
\end{equation}
where $||A||_1 = \mbox{Tr} \sqrt{A^{\dagger} A}$ denotes the trace norm of $A$. However, the computation of the trace norm in Eq. (\ref{error-2}) becomes incredibly tedious for large $M$. Also, it is difficult to imagine the large $M$ behavior of
the minimal error probability from Eq. (\ref{error-2}). In order to overcome these difficulties the quantum Chernoff (QC) bound was considered\cite{chernoff-1,chernoff-2}. The QC bound $P_{QC}$ between $\rho_0$ and $\rho_1$ is defined as 
\begin{equation}
\label{chernoff-1}
P_{QC} = \frac{1}{2} \left(\min_{s \in [0,1]} Q_s \right)^M
\end{equation}
where 
\begin{equation}
\label{QC-11}
Q_s =  \mbox{Tr} \left[\rho_0^s \rho_1^{1-s} \right].
\end{equation}
This gives a tight upper bound for $P_{E}^{min}$, i.e., $P_{E}^{min} \leq P_{QC}$. This bound has been analytically computed in several simple quantum systems \cite{chernoff-2}. However, the computation of the 
optimal value $s_*$, which minimizes $Q_s$, is in general highly tedious. 
Therefore, in Ref.\cite{tan08} the quantum  Bhattacharyya (QB) bound $P_{QB}$ between $\rho_0$ and $\rho_1$ was computed, where $s = 1 / 2$ is chosen instead of the optimal value $s = s_*$. For this reason, 
$P_{QB}$ is always larger than $P_{QC}$ if $s_*$ is not $1/2$.
If $N_S \ll 1 \ll N_B$,  the final form of the QB bound between $\rho_0$ and $\rho_1$ reduces to
\begin{equation}
\label{QB-2-1}
P_{QB} \approx \frac{1}{2} \exp \left[ - \frac{M}{4 N_B} \frac{\kappa C^2}{A + \sqrt{A^2 - 1}} \right]  \approx \frac{1}{2} \exp \left[ - \frac{M \kappa N_S}{N_B} \right].
\end{equation}
For classical coherent-state  illumination the corresponding QB bound\footnote{In this case $s_* = 1/2$ and hence $P_{QC} = P_{QB}$.} is 
\begin{equation}
\label{QB-1-2}
P_{QB}^{(1)} = \frac{1}{2} \exp \left[ - \frac{\sqrt{1 + N_B} - \sqrt{N_B}}{\sqrt{1 + N_B} + \sqrt{N_B}} M \kappa N_S \right] \approx \frac{1}{2} \exp \left[ - \frac{M \kappa N_S}{4 N_B} \right].
\end{equation}
The difference between Eq. (\ref{QB-2-1}) and Eq. (\ref{QB-1-2}) is a missing of factor $4$ in the exponent of Eq. (\ref{QB-2-1}), which implies the quantum advantage  of $6$ dB when
comparing classical illumination with this approach.

\section{Two single-mode squeezing operations}

The single-mode squeezing operation  is defined as 
\begin{equation}
\label{singleSO}
\hat{S} (z) = \exp \left[ \frac{1}{2} \left( z^* \hat{a}^2 - z \hat{a}^{\dagger 2} \right) \right]
\end{equation}
where $z = r e^{i \phi}$ and, $\hat{a}$ and $\hat{a}^{\dagger}$ are creation and annihilation operators, respectively. Then, two single-mode squeezing operations can be written as 
\begin{equation}
\label{twoSO}
\hat{S} (z_1, z_2) = \exp \left[ \frac{1}{2} \left( z_1^* \hat{a}_1^2 - z_1 \hat{a}_1^{\dagger 2} \right) \right] \exp \left[ \frac{1}{2} \left( z_2^* \hat{a}_2^2 - z_2 \hat{a}_2^{\dagger 2} \right) \right] = \exp \left[{\frac{1}{2} \hat{r}^T \bar{H}_1 \hat{r}}\right]
\end{equation}
where $z_i = r_i e^{i \phi_i}$, $\hat{r} = (\hat{x}_1, \hat{p}_1, \hat{x}_2, \hat{p}_2)^T$, and 
\begin{eqnarray}
\label{def-H-1}
\bar{H}_1 = \left(          \begin{array}{cc}
                  -r_1 \sin \phi_1  &   r_1 \cos \phi_1          \\
                   r_1 \cos \phi_1  &  r_1  \sin \phi_1           
                       \end{array}                                         \right)         \oplus
    \left(          \begin{array}{cc}
                  -r_2 \sin \phi_2  &   r_2 \cos \phi_2          \\
                   r_2 \cos \phi_2  &  r_2  \sin \phi_2           
                       \end{array}                                         \right).
\end{eqnarray}
The direct sum $\oplus$ acts on two matrices $A$ and $B$ such that $A \oplus B = \left( \begin{array}{cc}  A & 0     \\    0  & B   \end{array}  \right)$.
Now, we choose $\phi_1 = \phi_2 = 0$ for simplicity\footnote{In Ref. \cite{Jo21-1} it was shown that $\phi_1 = \phi_2$ is an optimal condition under a realistic receiver.}, which makes $\bar{H}_1$ to be $\bar{H}_1 = (r_1 \sigma_x) \oplus (r_2 \sigma_x)$, where $\sigma_x$ is the $x$-component 
of the Pauli matrices. 

The symplectic transform matrix $M_S$ corresponding to $\hat{S} (r_1, r_2)$ is 
\begin{eqnarray}
\label{symplectic-1}
M_S &=& e^{\Omega \bar{H}_1}                                                                                \\     \nonumber
&=& \mbox{diag} (\cosh r_1 + \sinh r_1, \cosh r_1 - \sinh r_1, \cosh r_2 + \sinh r_2, \cosh r_2 - \sinh r_2)
\end{eqnarray}
where $\Omega = -i [\hat{r}, \hat{r}^T] = \left(  \begin{array}{cc} 0 & 1  \\   -1 & 0  \end{array}  \right)   \oplus \left(  \begin{array}{cc} 0 & 1  \\   -1 & 0  \end{array}  \right) $.
Now, we define $n_j \equiv \sinh^2 r_j$, which is the mean photon number of a squeezed vacuum state\cite{weedbrook12}. Then, $M_S$ can be written as
\begin{equation}
\label{symplectic-2}
M_S = \mbox{diag} (\gamma_{1,+}, \gamma_{1,-}, \gamma_{2,+}, \gamma_{2,-} )
\end{equation}
where $\gamma_{j,\pm} = \sqrt{n_j + 1} \pm \sqrt{n_j}$.

The operation  $\hat{S} (r_1, r_2)$ on the TMSV state changes the covariance matrix to 
\begin{eqnarray}
\label{CM-ASTM-1}
V_{TSS} = M_S V_{TMSV} M_S^T = \left(            \begin{array}{cccc}
                                                                   A \gamma_{1,+}^2   &   0   &   C \gamma_{1,+}  \gamma_{2,+}   &   0                                 \\
                                                                          0   &   A \gamma_{1,-}^2   &   0   &   -C \gamma_{1,-} \gamma_{2,-}                          \\
                                                                  C \gamma_{1,+} \gamma_{2,+}   &   0   &   A \gamma_{2,+}^2   &   0                                    \\
                                                                  0   &   -C \gamma_{1,-} \gamma_{2,-}   &   0   &   A \gamma_{2,-}^2  
                                                                            \end{array}                                                                                \right)
\end{eqnarray}
where the subscript TSS stands for ``two single-mode squeezing''. It is worthwhile noting that Eq. (\ref{CM-ASTM-1}) implies that the average photon numbers per S and I modes are 
\begin{equation}
\label{error-1}
\tilde{N}_S = N_S + 2 n_1 N_S + n_1      \hspace{1.0cm}  \tilde{N}_I = N_S + 2 n_2 N_S + n_2.
\end{equation}
Thus, they are different from each other if $r_1 \neq r_2$. Another point we have to note from Eq. (\ref{CM-ASTM-1}) is the fact that the entanglement of the TSS state is exactly the same with that of the TMSV state, because 
the operation $\hat{S} (z_1, z_2)$ is merely a local unitary operation. This fact can be proved explicitly by computing  the logarithmic negativities $E_{{\cal N}}$, which yields
\begin{equation}
\label{entanglement-1}
E_{{\cal N}} (TMSV) = E_{{\cal N}} (TSS) = -2 \log_2 (\sqrt{1 + N_S} - \sqrt{N_S}).
\end{equation}

The corresponding states $\rho_0$ for the null hypothesis $H_0$ and $\rho_1$ for the alternative hypothesis $H_1$ become zero-mean Gaussian states with 
covariance matrices 
\begin{eqnarray}
\label{CM-rho0}
V_0 = \left(                   \begin{array}{cccc}
                         B   &   0   &   0   &   0                                            \\
                         0   &   B   &   0   &   0                                            \\
                         0   &   0   &   A \gamma_{2,+}^2   &   0                  \\ 
                         0   &   0   &   0   &   A \gamma_{2,-}^2
                                   \end{array}                                                        \right)
\end{eqnarray}
for $\rho_0$  and
\begin{eqnarray}
\label{CM-rho1}
V_1 =  \left(                    \begin{array}{cccc}
                       F_+   &   0   &   \sqrt{\kappa} C \gamma_{1,+} \gamma_{2,+}   &   0                                    \\
                       0   &   F_-   &   0   &   - \sqrt{\kappa} C \gamma_{1,-} \gamma_{2,-}                                 \\
                       \sqrt{\kappa} C \gamma_{1,+} \gamma_{2,+}   &   0   &   A \gamma_{2,+}^2   &   0               \\
                       0   &    - \sqrt{\kappa} C \gamma_{1,-} \gamma_{2,-}   &   0   &    A \gamma_{2,-}^2
                                     \end{array}                                                                                                       \right)
\end{eqnarray}
for $\rho_1$, where $F_{\pm} = B + \kappa (A \gamma_{1,\pm}^2 - 1)$. 
It is straightforward to show that $V_0$ and $V_1$ reduce to the corresponding covariance matrices (\ref{TMSV-0}) and (\ref{TMSV-1}) respectively when $\gamma_{j,\pm} = 1$. 
It is easy to show $\lim_{\kappa \rightarrow 0} V_1 = V_0$.

The matrix $V_0$ in Eq. (\ref{CM-rho0}) can be expressed as 
\begin{eqnarray}
\label{symplectic-3}
V_0 = S_{V0}  \left(    \begin{array}{cc}
                                                \alpha_1 \openone_2   &   0   \\
                                                 0   &   \alpha_2   \openone_2
                                                   \end{array}                       \right)   S_{V0}^T
\end{eqnarray}
where $\openone_2$ is $2 \times 2$ identity matrix and, $\alpha_1 = B$, $\alpha_2 = A$, and $S_{V0} = \mbox{diag} (1, 1, \zeta^{-1}, \zeta)$ with $\zeta = \sqrt{\gamma_{2,-} / \gamma_{2,+}}$.
When deriving the symplectic eigenvalue $\alpha_2$ we used $\gamma_{2,+} \gamma_{2,-} = 1$ explicitly. The matrix $V_1$ in Eq. (\ref{CM-rho1}) also can be expressed as 
\begin{eqnarray}
\label{symplectic-4}
V_1 = S_{V1}  \left(    \begin{array}{cc}
                                                \beta_1 \openone_2   &   0   \\
                                                 0   &   \beta_2   \openone_2
                                                   \end{array}                       \right)   S_{V1}^T.
\end{eqnarray}
The symplectic eigenvalues $\beta_1$ and $\beta_2$ are 
\begin{equation}
\label{symplectic-beta-1}
\beta_1 = \sqrt{\frac{G + 2 H + \xi}{2}}    \hspace{1.0cm} \beta_2 = \sqrt{\frac{G + 2 H - \xi}{2}}
\end{equation}
where $\xi = \sqrt{G^2 - 4 \kappa C^2 G_+ G_-}$ and 
\begin{eqnarray}
\label{boso-1}
&&G = F_+ F_- - A^2      \hspace{1.0cm}                  H = A^2 - \kappa C^2                                      \\    \nonumber
&&G_{\pm} = F_{\pm} \gamma_{1,\mp} - A \gamma_{1,\pm}    \hspace{1.0cm}   H_{\pm} = A F_{\pm} - \kappa C^2 \gamma_{1,\pm}^2.
\end{eqnarray}
It is worthwhile noting that $G$, $H$, $G_{\pm}$, and $H_{\pm}$ are independent of $\gamma_{2,\pm}$ because they are decoupled due to the identity $\gamma_{2,+} \gamma_{2,-} = 1$.
The symplectic transform $S_{V1}$ becomes
\begin{eqnarray}
\label{symplectic-5}
S_{V1} =  \left(                \begin{array}{cccc}
                                y_1   &   0   &   y_5   &   0                  \\
                                0   &   y_2   &   0   &   y_6                  \\
                                y'_5  &   0   &   y_3   &   0                \\
                                0   &   y'_6  &   0   &   y_4
                                      \end{array}                                   \right)
\end{eqnarray}
where
\begin{eqnarray}
\label{boso-2}
&&y_1 = \frac{\kappa C^2 G_+^2 H_+}{\sqrt{\beta_1 \xi \Delta_1} \Delta_2}                  \hspace{1.0cm}   y_2 = \frac{1}{2} \sqrt{\frac{\beta_1}{\xi \Delta_1}} \frac{\kappa C^2 G_+}{\Delta_2}  \left[ 2 A G_+ - \gamma_{1,+} (G - \xi) \right]  \\    \nonumber
&&y_3 = \frac{1}{2} \frac{\sqrt{\kappa} C G_+ H_+ (G + \xi)}{\sqrt{\beta_2 \xi \Delta_2} \Delta_1} \gamma_{2,+}              \hspace{0.5cm}
y_4 = \frac{1}{2} \sqrt{\frac{\beta_2}{\xi \Delta_2}} \frac{\sqrt{\kappa} C G_+}{\Delta_1} \gamma_{2,-} \left[ F_+ (G + \xi) - 2 \kappa C^2 G_+ \gamma_{1,+} \right]                         \\                   \nonumber
&&y_5 = \frac{\kappa C^2 G_+^2 H_+}{\sqrt{\beta_2 \xi \Delta_2} \Delta_1}     \hspace{1.0cm}  y_6 = - \frac{1}{2} \sqrt{\frac{\beta_2}{\xi \Delta_2}} \frac{\kappa C^2 G_+}{\Delta_1} \left[ \gamma_{1,+} (G + \xi) - 2 A G_+ \right]   \\   \nonumber
&&y'_5 = \frac{1}{2} \frac{\sqrt{\kappa} C G_+ H_+ (G - \xi)}{\sqrt{\beta_1 \xi \Delta_1} \Delta_2} \gamma_{2,+}      \hspace{0.5cm}
y'_6 = -\frac{1}{2} \sqrt{\frac{\beta_1}{\xi \Delta_1}} \frac{\sqrt{\kappa} C G_+}{\Delta_2} \gamma_{2,-} \left[ 2 \kappa C^2 G_+ \gamma_{1,+} - F_+ (G - \xi) \right]
\end{eqnarray}
with $\Delta_1 = F_+ \beta_1^2 - A H_+$ and $\Delta_2 = A H_+ - F_+ \beta_2^2$. 
When $\gamma_{1,\pm} = \gamma_{2,\pm} = 1$, it is straightforward to show that $y_1$, $y_2$, $y_3$, and $y_4$ reduce to $x_+$ and, $y_5$, $y'_5$, $-y_6$, and $-y'_6$ to $x_-$ 
where $x_{\pm} = \sqrt{\frac{(1 + \kappa) A - \kappa + B \pm (\beta_1 + \beta_2)}{2 (\beta_1 + \beta_2)}}$. This is consistent with the results in Ref.\cite{tan08}.

In order to compute $Q_s$ between $\rho_0$ and $\rho_1$ given by Eq. (\ref{QC-11}), we define 
\begin{equation}
\label{boso-3}
\Lambda_p (x) = \frac{(x + 1)^p + (x - 1)^p}{(x + 1)^p - (x - 1)^p}          \hspace{1.0cm}   G_p (x) = \frac{2^p}{(x + 1)^p - (x - 1)^p}.
\end{equation}
Also we define 
\begin{eqnarray}
\label{boso-4}
&&\Sigma_0 (s) = S_{V0}  \left(    \begin{array}{cc}
                                                \Lambda_s(\alpha_1) \openone_2   &   0   \\
                                                 0   &   \Lambda_s (\alpha_2)   \openone_2
                                                   \end{array}                       \right)   S_{V0}^T                      \\   \nonumber
&&  \Sigma_1 (1 - s) = S_{V1}  \left(    \begin{array}{cc}
                                                \Lambda_{1 - s} (\beta_1) \openone_2   &   0   \\
                                                 0   &   \Lambda_{1 - s} (\beta_2)   \openone_2
                                                   \end{array}                       \right)   S_{V1}^T.
\end{eqnarray}
Then, for the case of general $n$-mode Gaussian states $\rho_0$ and $\rho_1$  $Q_s$ becomes\cite{computable-1}
\begin{equation}
\label{gaus-cher-2}
Q_s = \bar{Q}_s \exp \bigg[ -(\bar{x}_0 - \bar{x}_1)^T \Sigma^{-1} (s)  (\bar{x}_0 - \bar{x}_1) \bigg]
\end{equation}
where $\Sigma (s) = \Sigma_0 (s) + \Sigma_1 (1 - s)$ and 
\begin{equation}
\label{gaus-cher-3}
\bar{Q}_s = \frac{2^n \prod_{k=1}^n G_s (\alpha_k) G_{1 - s} (\beta_k)}{\sqrt{\det \Sigma (s)}}.
\end{equation}
In Eq. (\ref{gaus-cher-2}) $\bar{x}_0$ and $\bar{x}_1$ are the mean vector of $\rho_0$ and $\rho_1$. 

Since $\bar{x}_0 = \bar{x}_1 = 0$ and $n = 2$ for our case, $Q_s$ becomes 
\begin{equation}
\label{gaus-cher-4}
Q_s = \bar{Q}_s = \frac{4 G_s (\alpha_1) G_s (\alpha_2) G_{1 - s} (\beta_1) G_{1 - s} (\beta_2)}{\sqrt{\det \Sigma (s)}}.
\end{equation}
It is straightforward to show
\begin{equation}
\label{det-1}
\det \Sigma (s) = \left[x_1 (s) x_3 (s) - x_5^2 (s) \right] \left[ x_2 (s) x_4 (s) - x_6^2 (s) \right]
\end{equation}
where
\begin{eqnarray}
\label{det-2}
&&x_1 (s) = \Lambda_s (\alpha_1) + y_1^2 \Lambda_{1 - s} (\beta_1) + y_5^2 \Lambda_{1 - s} (\beta_2)    \\   \nonumber
&&x_2 (s) =  \Lambda_s (\alpha_1) + y_2^2 \Lambda_{1 - s} (\beta_1) + y_6^2 \Lambda_{1 - s} (\beta_2)    \\   \nonumber
&&x_3 (s) = \zeta^{-2}  \Lambda_s (\alpha_2) + {y'_5}^2 \Lambda_{1 - s} (\beta_1) + y_3^2 \Lambda_{1 - s} (\beta_2)    \\   \nonumber
&&x_4 (s) = \zeta^{2}  \Lambda_s (\alpha_2) + {y'_6}^2 \Lambda_{1 - s} (\beta_1) + y_4^2 \Lambda_{1 - s} (\beta_2)     \\   \nonumber
&&x_5 (s) = y_1 y'_5 \Lambda_{1 - s} (\beta_1) + y_3 y_5 \Lambda_{1 - s} (\beta_2)                                      \\   \nonumber
&&x_6 (s) = y_2 y'_6 \Lambda_{1 - s} (\beta_1) + y_4 y_6 \Lambda_{1 - s} (\beta_2).                                  
\end{eqnarray}
Therefore, the QC bound can be computed by inserting Eqs. (\ref{gaus-cher-4}) and (\ref{det-1}) into Eq. (\ref{chernoff-1}) to obtain the optimal value $s_*$. 
In order to compute $s_*$ we should solve 
\begin{equation}
\label{optimal}
\frac{d Q_s}{d s} \Bigg|_{s = s_*} = 0.
\end{equation}
However, it seems to be impossible to solve Eq. (\ref{optimal}) analytically. Thus, in this paper, instead of finding the QC bound, we will compute
the QB bound, which is defined as 
\begin{equation}
\label{gaus-cher-5}
P_{QB} = \frac{1}{2} \left(  Q_{s=1/2} \right)^M.
\end{equation}

\begin{figure}[ht!]
\begin{center}
\includegraphics[height=5.0cm]{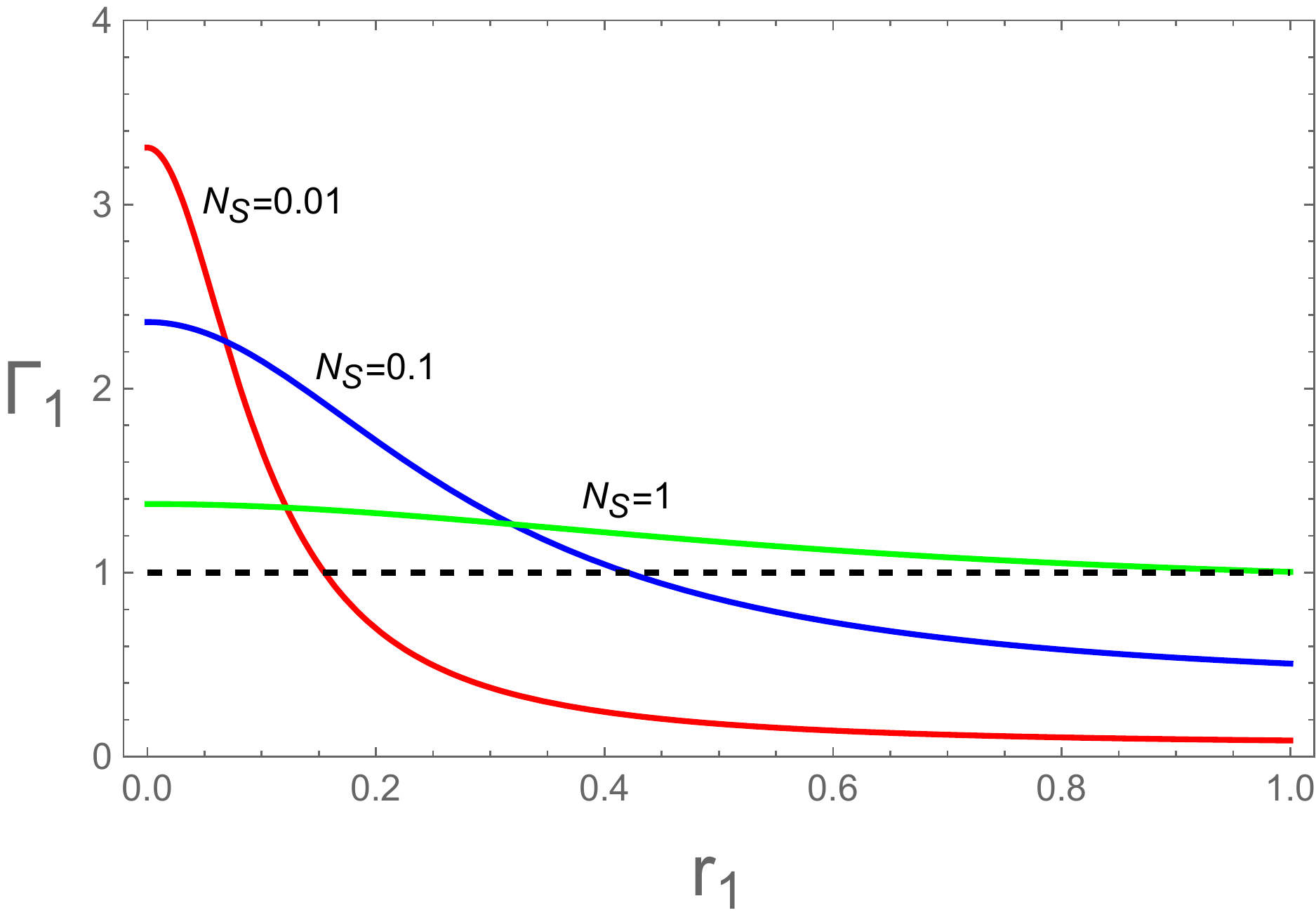} 
\includegraphics[height=5.0cm]{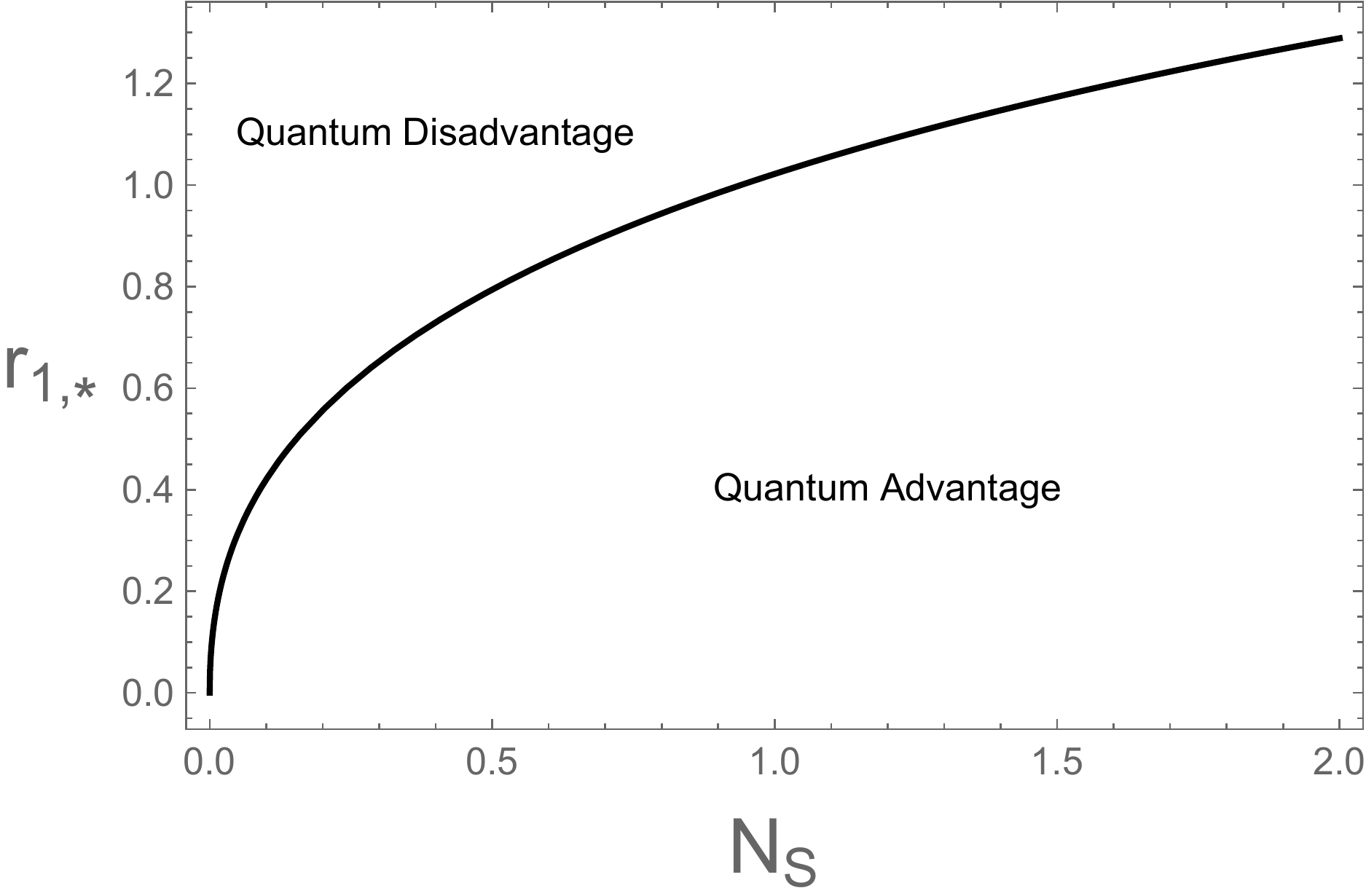}

\caption[fig1]{(Color online) (a) The $r_1$-dependence of $\Gamma_1$ when $N_S = 0.01$, $0.1$, and $1$. This figure shows that $\Gamma_1$ decreases with increasing $r_1$, and eventually $\Gamma_1$ becomes 
less than $1$ when $r_1 > r_{1,*}$. This means that the quantum disadvantage occurs in this region. This figure with Eq. (\ref{entanglement-1}) implies that entanglement is not the only resource responsible for the quantum advantage in quantum illumination.
(b) The $N_S$-dependence of $r_{1,*}$.
 }
\end{center}
\end{figure}

Now, we assume $N_B \gg 1$ and $N_B \gg N_S$ with $\kappa \ll 1$. 
In this case, after a long calculation one can show
\begin{eqnarray}
\label{approx-1}
&&4 G_{1/2} (\alpha_1)  G_{1/2} (\alpha_2)  G_{1/2} (\beta_1)   G_{1/2} (\beta_2) \approx 16 N_B (A + \sqrt{A^2 - 1}) \left(1 + \frac{K_1}{8 N_B} \right)  \\   \nonumber
&&\sqrt{\det \Sigma (1 / 2)} \approx 16 N_B (A + \sqrt{A^2 - 1}) \left(1 + \frac{K_2}{8 N_B} \right) 
\end{eqnarray} 
where
\begin{eqnarray}
\label{approx-2}
&& K_1 = 2 (2 - \kappa) + \kappa A (\gamma_{1,+}^2 + \gamma_{1,-}^2) - \frac{\kappa C^2}{\sqrt{A^2 - 1}}  (\gamma_{1,+}^2 + \gamma_{1,-}^2)   \\   \nonumber
&&K_2 =  2 (2 - \kappa) + \kappa A (\gamma_{1,+}^2 + \gamma_{1,-}^2) - \frac{\kappa A C^2}{\sqrt{A^2 - 1} (A + \sqrt{A^2 - 1})}  (\gamma_{1,+}^2 + \gamma_{1,-}^2).
\end{eqnarray}
Therefore, $P_{QB}$ can be written approximately as
\begin{equation}
\label{approx-3}
P_{QB} \approx \frac{1}{2} \exp \left[ - \frac{M \kappa C^2}{4 N_B} \frac{2 n_1 + 1}{A + \sqrt{A^2 - 1}} \right].
\end{equation}
It is worthwhile noting that $P_{QB}$ in Eq. (\ref{approx-3}) is independent of $\gamma_{2,\pm}$. This is due to the fact that the constraint $\gamma_{2,+} \gamma_{2,-} = 1$ decouples $\gamma_{2,+}$ and $\gamma_{2,-}$  in $K_1$ and $K_2$.
Thus, the squeezing operation of the second single-mode does not change the QB bound.
Compared Eq. (\ref{approx-3}) to a classical coherent-state  illumination (\ref{QB-1-2}) with changing $N_S \rightarrow \tilde{N}_S$,
one can show that a quantum advantage of our case is $10 \log_{10} \Gamma_1$ (dB), where
\begin{equation}
\label{error-3}
\Gamma_1 = \frac{4 N_S (1 + N_S) (2 n_1 + 1)}{(N_S + 2 n_1 N_S + n_1) (\sqrt{1 + N_S} + \sqrt{N_S})^2} = \frac{4 (\tilde{N}_S - n_1) (\tilde{N}_S + n_1 + 1)}{\tilde{N}_S (\sqrt{\tilde{N}_S + n_1 + 1} + \sqrt{\tilde{N}_S - n_1} )^2}.
\end{equation}
When $n_1 = 0$ and $N_S \ll 1$, it is easy to show $\Gamma_1 \approx 4$ as expected. The $r_1$-dependence of $\Gamma_1$ is plotted in Fig. 1(a) for various $N_S$. This figure shows that $\Gamma_1$ decreases with increasing $r_1$.
This can be expected because the TMSV state is nearly optimal\cite{palma18,nair20,brad21} in the error probability for continuous-variable quantum illumination. When $r_1$ approaches to $\infty$, $\Gamma_1 - 1$ approaches to some negative value.
This means that the quantum advantage disappears when $r_1 \geq r_{1,*}$. The $N_S$-dependence of $r_{1,*}$ is plotted in Fig. 1(b). As Fig. 1(a) shows, $r_{1,*}$ exhibits a monotonic behavior.


\section{Effect of two-mode squeezing operation}

In this section we examine the effects of the two-mode squeezing operation on two-mode Gaussian quantum illumination.
The two-mode squeezing operation is defined as 
\begin{equation}
\label{two-mode-SO}
\hat{S}_{2} (z) = \exp \left[ z^* \hat{a}_1 \hat{a}_2 - z \hat{a}_1^{\dagger} \hat{a}_2^{\dagger} \right] \equiv e^{\frac{i}{2} \hat{r}^T \bar{H}_{2} \hat{r} }
\end{equation}
where $z = r e^{i \phi}$, $\hat{r} = (\hat{x}_1, \hat{p}_1, \hat{x}_2, \hat{p}_2)^T$, and 
\begin{eqnarray}
\label{def-H-2}
\bar{H}_2 = \left(                           \begin{array}{cc}
                              0  &  r (\sigma_x \cos \phi  - \sigma_z \sin \phi )                      \\
                               r (\sigma_x \cos \phi  - \sigma_z \sin \phi )  &  0
                                                      \end{array}                                           \right).
\end{eqnarray}
Let $\phi = 0$ for simplicity. Then, the corresponding symplectic transformation matrix is 
\begin{eqnarray}
\label{symplectic-11}
S_2 = e^{\Omega \bar{H}_2} = \left(                        \begin{array}{cc}
                                                              \openone_2 \cosh r  &  \sigma_z \sinh r         \\
                                                              \sigma_z \sinh r  &  \openone_2 \cosh r
                                                                               \end{array}                               \right).
\end{eqnarray}    
The $\hat{S}_2 (r)$ operation on the TMSV state means the covariance matrix to becomes 
\begin{eqnarray}
\label{CM-TMS-1}
V_{TMS} = S_2 V_{TMAS} S_2^T = \left(                                  \begin{array}{cccc}
                                                            \tilde{A}  &  0  &  \tilde{C}  &  0                           \\
                                                            0  &  \tilde{A}  &  0  &  -\tilde{C}                          \\
                                                            \tilde{C}  &  0  &  \tilde{A}  &  0                            \\
                                                            0  &  -\tilde{C}  &  0  &  \tilde{A}
                                                                                             \end{array}                                  \right)
\end{eqnarray}
where  the subscript TMS stands for ``two-mode squeezing'' and 
\begin{equation}
\label{CM-TMS-2}
\tilde{A} = A \cosh 2 r + C \sinh 2 r   \hspace{1.0cm}  \tilde{C} = A \sinh 2 r + C \cosh 2 r.
\end{equation}   
Eq. (\ref{CM-TMS-1}) implies that the average photon numbers per S and I modes are 
\begin{equation}
\label{error-4-1}
\bar{N}_S = \bar{N}_I = \frac{1}{2} (\tilde{A} - 1 ).
\end{equation}
Another point we want to note is the fact that entanglement of the TMS state can be different from that of the TMSV state, because $\hat{S}_{2} (z)$ is a global unitary operator.
In fact, the logarithmic negativity of the TMS state is 
\begin{equation}
\label{entanglement-2}
E_{{\cal N}} (TMS) - E_{{\cal N}} (TMSV) = 2 r \log_2 e = 2.89 r \geq 0.
\end{equation}
Therefore, the entanglement  of TMS state is larger than that of TMSV state.

Using Eq. (\ref{CM-TMS-1}) one can explicitly derive the states $\rho_0$ and $\rho_1$ for the null and alternative hypotheses. 
Both of them are zero-mean Gaussian states whose covariance matrices are
\begin{eqnarray}
\label{TMS-rho0}
V_0 = \left(                 \begin{array}{cccc}
                B  &  0  &  0  &  0                        \\
                0  &  B  &  0  &  0                        \\
                0  &  0  &  \tilde{A}  &  0            \\
                0  &  0  &  0  &  \tilde{A}
                                 \end{array}                               \right) \equiv  S_{V0}  \left(    \begin{array}{cc}
                                                \alpha_1 \openone_2   &   0   \\
                                                 0   &   \alpha_2   \openone_2
                                                   \end{array}                       \right)   S_{V0}^T
\end{eqnarray}
for $\rho_0$  and 
\begin{eqnarray}
\label{TMS-rho1}
V_1 = \left(           \begin{array}{cccc}
                     \tilde{F}  &  0  &  \sqrt{\kappa} \tilde{C}  &  0               \\
                      0  &  \tilde{F}  &  0  &  -\sqrt{\kappa} \tilde{C}              \\
                      \sqrt{\kappa} \tilde{C}  &  0  &  \tilde{A}  &  0    \\
                      0  &  -\sqrt{\kappa} \tilde{C}  &  0  &  \tilde{A}
                          \end{array}                                                   \right)  \equiv  S_{V1}  \left(    \begin{array}{cc}
                                                \beta_1 \openone_2   &   0   \\
                                                 0   &   \beta_2   \openone_2
                                                   \end{array}                       \right)   S_{V1}^T
\end{eqnarray}
for $\rho_1$ with $\tilde{F} = \kappa \tilde{A} + B - \kappa$. One can show $\lim_{\kappa \rightarrow 0} V_1 = V_0$. 
The symplectic eigenvalues $\alpha_j$ and $\beta_j$ are 
\begin{equation}
\label{symp-eigen}
\alpha_1 = B   \hspace{.5cm}  \alpha_2 = \tilde{A}  \hspace{.5cm} \beta_k = \frac{1}{2} \left[ (-1)^k (\tilde{A} - \tilde{F}) + \sqrt{(\tilde{F} + \tilde{A})^2 - 4 \kappa \tilde{C}^2 } \right].
\end{equation}   
Also one can show that the symplectic transformation matrices are $ S_{V0} = \openone_4$ and    
\begin{eqnarray}
\label{symplectic-12}
S_{V1} =  \left(                  \begin{array}{cc}
                                         X_+  &  X_-             \\
                                         X_-  &  X_+
                                         \end{array}                                   \right)
\end{eqnarray}         
where $X_{\pm} = \mbox{diag} (x_{\pm}, \pm x_{\pm} )$ with
\begin{equation}
\label{symplectic-13}
x_{\pm} = \sqrt{\frac{\tilde{F} + \tilde{A} \pm (\beta_1 + \beta_2)}{2 (\beta_1 + \beta_2)}}.
\end{equation}

Using Eqs. (\ref{chernoff-1}),  (\ref{boso-4}), (\ref{gaus-cher-2}), and (\ref{gaus-cher-4}), one can show that the QC bound is 
\begin{equation}
\label{chernoff-11}
P_{QC} = \frac{1}{2} \left( \min_{s \in [0, 1]} \frac{4 G_s (\alpha_1) G_s (\alpha_2) G_{1 - s} (\beta_1) G_{1 - s} (\beta_2)}{y_1 (s) y_2 (s) - z_3 (s)^2} \right)^M
\end{equation}
where 
\begin{eqnarray}
\label{chernoff-12}
&& y_1 (s) = \Lambda_{1 - s} (\beta_1) x_+^2 + \Lambda_{1 - s} (\beta_2) x_-^2 + \Lambda_s (\alpha_1)       \\    \nonumber
&& y_2 (s) = \Lambda_{1 - s} (\beta_1) x_-^2 + \Lambda_{1 - s} (\beta_2) x_+^2 + \Lambda_s (\alpha_2)       \\    \nonumber
&& z_3 (s) = \left( \Lambda_{1 - s} (\beta_1) + \Lambda_{1 - s} (\beta_2)  \right) x_+ x_-.
\end{eqnarray}

\begin{figure}[ht!]
\begin{center}
\includegraphics[height=6.0cm]{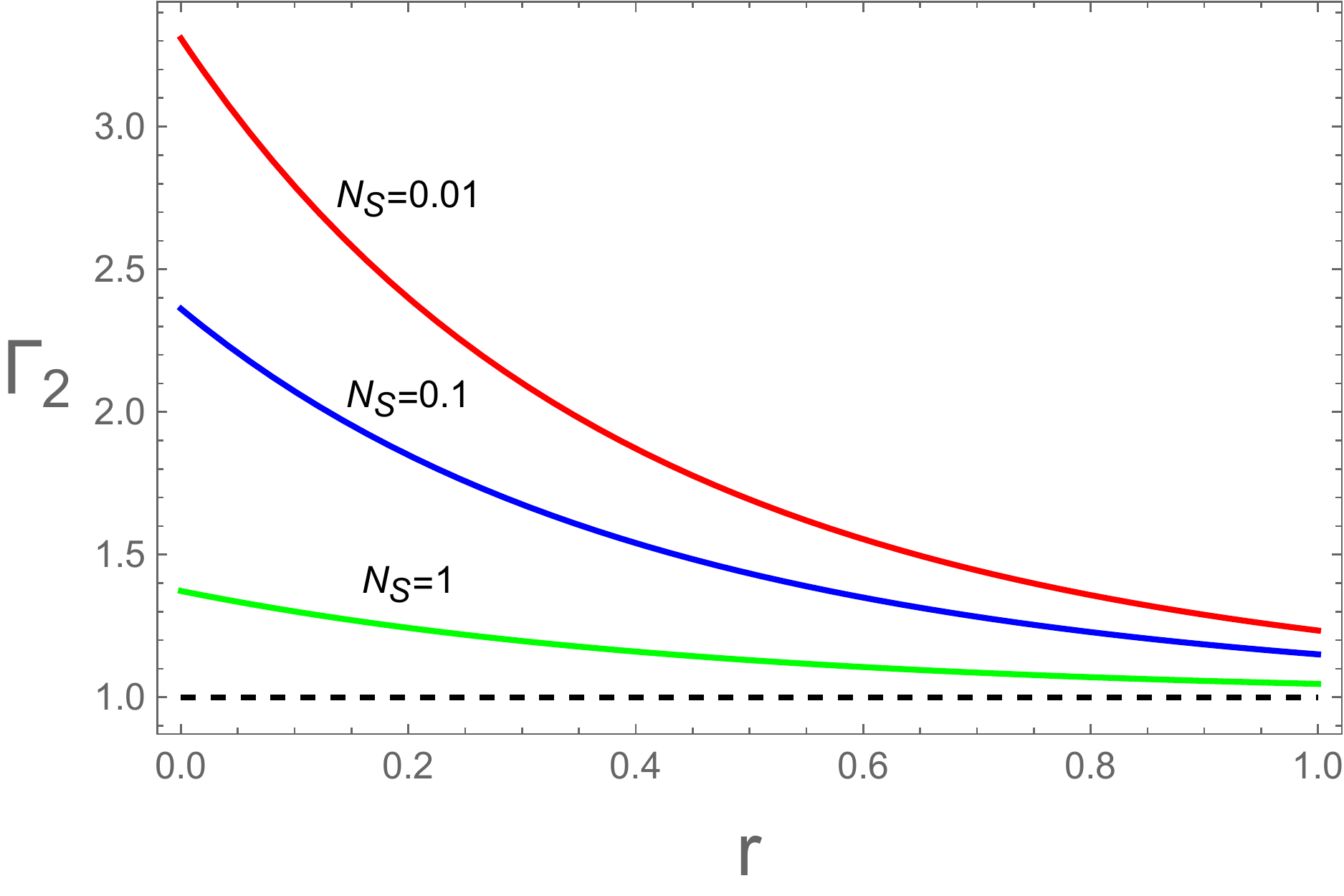} 

\caption[fig2]{(Color online) The $r$-dependence of $\Gamma_2$ when  $N_S = 0.01$, $0.1$, and $1$. This figure shows that $\Gamma_2$ decreases with increasing $r$.
This figure with Eq. (\ref{entanglement-2}) implies that  entanglement is not the only resource responsible for the quantum advantage in quantum illumination.
 }
\end{center}
\end{figure}

Now, let us compute the QB bound. From (\ref{chernoff-11}) the QB bound becomes 
\begin{equation}
\label{QB-11}
P_{QB} = \frac{1}{2} \left(\frac{4 G_{1/2} (\alpha_1) G_{1/2} (\alpha_2) G_{1/2} (\beta_1) G_{1/2} (\beta_2)}{y_1 (1/2) y2 (1/2) - z_3 (1/2)^2} \right)^M.
\end{equation} 
If we assume $N_B \gg N_S$ and $N_B \gg 1$ with $\kappa \ll 1$, one can show 
\begin{eqnarray}
\label{approx-11}
&&4 G_{1/2} (\alpha_1) G_{1/2} (\alpha_2) G_{1/2} (\beta_1) G_{1/2} (\beta_2) \approx 16 N_B (\tilde{A} + \sqrt{\tilde{A}^2 - 1}) \left( 1 + \frac{J_1}{4 N_B} \right)   \\   \nonumber
&&y_1 (1/2) y2 (1/2) - z_3 (1/2)^2 \approx  16 N_B (\tilde{A} + \sqrt{\tilde{A}^2 - 1}) \left( 1 + \frac{J_2}{4 N_B} \right) 
\end{eqnarray}
where
\begin{equation}
\label{approx-12}
J_1 = (2 - \kappa + \kappa \tilde{A}) - \frac{\kappa \tilde{C}^2}{\sqrt{\tilde{A}^2 - 1}}                  \hspace{.5cm}
J_2 =  (2 - \kappa + \kappa \tilde{A}) - \frac{\kappa \tilde{C}^2 \tilde{A}}{\sqrt{\tilde{A}^2 - 1} (\tilde{A} + \sqrt{\tilde{A}^2 - 1})}.
\end{equation}
Therefore, the QB bound $P_{QB}$ becomes approximately 
\begin{equation}
\label{approx-13}
P_{QB} \approx \frac{1}{2} \exp \left[ - \frac{M}{4 N_B} \frac{\kappa \tilde{C}^2}{\tilde{A} + \sqrt{\tilde{A}^2 - 1}} \right].
\end{equation}
Compared Eq. (\ref{approx-13}) to a classical coherent-state illumination (\ref{QB-1-2}) with changing $N_S \rightarrow \bar{N}_S$, 
one can show that the quantum advantage is $10 \log_{10} \Gamma_2$ (dB), where
\begin{equation}
\label{error-4-2}
\Gamma_2 = \frac{\tilde{C}^2}{\bar{N}_S (\tilde{A} + \sqrt{\tilde{A}^2 - 1})}.
\end{equation}
The $r$-dependence of $\Gamma_2$ is plotted in Fig. 2 for various $N_S$. Like Fig. 1(a) $\Gamma_2$ decreases with increasing $r$. 
This figure shows that $\Gamma_2$ approaches to $1$ in the $r \rightarrow \infty$ limit regardless of $N_S$. Thus, quantum disadvantage does not occur in this case.



\section{Conclusion}
It is well-known that when $N_S \ll 1 \ll N_B$, quantum illumination with the TMSV state achieves a $6$ dB gain compared to classical coherent-state illumination even though original entanglement completely disappears at the final stage
due to strong background noise. It is believed that this quantum advantage is originated from the entanglement 
of the initial TMSV state. 
If this is right, it is very surprising because this fact implies that the benefits of entanglement can outlast entanglement itself.
Is this entanglement a unique resource for the $6$ dB gain? We try to give an answer in this paper to this question by introducing the squeezing operations.

First, we construct the TSS state by applying two single-mode squeezing operations to the TMSV state. It is obvious that the TSS state has the same entanglement with the 
TMSV state, because $\hat{S} (z_1, z_2)$ in Eq. (\ref{twoSO}) is merely a local unitary operator. Thus, if the entanglement is a unique origin of the $6$ dB gain, the same gain 
should be achieved in quantum illumination with the TSS state. However, as Fig. 1 shows, the quantum advantage reduces with increasing the squeezing parameter $r_1$, and 
eventually it disappears at $r_1 \geq r_{1,*}$. 

Second, we construct the TMS state by applying the two-mode squeezing operations to the TMSV state. It is shown that the TMS state has larger entanglement than the TMSV state. 
In spite of larger entanglement, Fig. 2 shows that the quantum advantage in quantum illumination with the TMS state reduces with increasing the squeezing parameter $r$. 

In fact, reduction of the quantum advantage can be expected, because it was proved in Ref. \cite{palma18,nair20,brad21} that the TMSV state is a nearly optimal state 
in the error probability provided that reflectivity is extremely small. From an aspect of entanglement reduction of the quantum advantage with increasing squeezing parameter 
implies that the initial entanglement is not the only resource for achieving the quantum advantage. Then, it is natural to ask a question: what are other resources, which are responsible for 
the quantum advantage of quantum illumination? Authors in Ref. \cite{discord1,discord2} have suggested that quantum discord is a genuine resource responsible for the quantum advantage. 
However, it was argued in Ref.\cite{anti-discord1} that the advantage can not be characterized by a quantum discord sorely. Further, the counterexample was found in Ref.\cite{Jo21-1}, which supports Ref.\cite{anti-discord1}.
In this reason, still we do not completely understand the genuine resource in quantum illumination. We hope to visit this issue in the future.

{\bf Acknowledgement}:
This work was supported by the National Research Foundation of Korea(NRF) grant funded by the Korea government(MSIT) (No. 2021R1A2C1094580).

\end{document}